\documentclass[9pt,conference]{IEEEtran}
\usepackage{amssymb,amsthm,amsmath,array}
\usepackage{graphicx}
\usepackage[caption=false,font=footnotesize]{subfig}
\usepackage{xspace}
\usepackage[sort&compress, numbers]{natbib}
\usepackage{stmaryrd}
\usepackage{xcolor}
\usepackage{mathtools}
\usepackage{float}
\usepackage{textcomp}

\begin{document}
\title{Unveiling the Hidden Agenda: Biases in News Reporting and Consumption}
\author{\IEEEauthorblockN{
        Alessandro Galeazzi\IEEEauthorrefmark{1}\IEEEauthorrefmark{4}, 
        Antonio Peruzzi\IEEEauthorrefmark{1}\IEEEauthorrefmark{4},  
        Emanuele Brugnoli\IEEEauthorrefmark{2},
        Marco Delmastro\IEEEauthorrefmark{1}\IEEEauthorrefmark{2}\IEEEauthorrefmark{5},
        Fabiana Zollo\IEEEauthorrefmark{1}\IEEEauthorrefmark{3}\IEEEauthorrefmark{5}
    }
    \IEEEauthorblockA{
        \IEEEauthorrefmark{1} Ca’ Foscari University of Venice, Italy\\
        \IEEEauthorrefmark{2} Centro Ricerche “Enrico Fermi” (CREF), Rome, Italy\\
        \IEEEauthorrefmark{3} The New Institute Centre for Environmental Humanities, Venice, Italy\\
        \IEEEauthorrefmark{4} A.G. and A.P. contributed equally to this work.\\
        \IEEEauthorrefmark{5} Correspondence to: {marco.delmastro, fabiana.zollo}@unive.it}
}
\maketitle
\begin{abstract}
    One of the most pressing challenges in the digital media landscape is understanding the impact of biases on the news sources that people rely on for information. Biased news can have significant and far-reaching consequences, influencing our perspectives and shaping the decisions we make, potentially endangering the public and individual well-being. With the advent of the Internet and social media, discussions have moved online, making it easier to disseminate both accurate and inaccurate information. To combat mis- and dis-information, many have begun to evaluate the reliability of news sources, but these assessments often only examine the validity of the news (narrative bias) and neglect other types of biases, such as the deliberate selection of events to favor certain perspectives (selection bias). This paper aims to investigate these biases in various news sources and their correlation with third-party evaluations of reliability, engagement, and online audiences. Using machine learning to classify content, we build a six-year dataset on the Italian vaccine debate and adopt a Bayesian latent space model to identify narrative and selection biases. Our results show that the source classification provided by third-party organizations closely follows the narrative bias dimension, while it is much less accurate in identifying the selection bias. Moreover, we found a nonlinear relationship between biases and engagement, with higher engagement for extreme positions. Lastly, analysis of news consumption on Twitter reveals common audiences among news outlets with similar ideological positions.
\end{abstract}

\section*{Introduction}
In the public sphere, diverse perspectives and opinions are frequently exchanged and debated, leading to the emergence of shared views and understanding~\cite{hab2010}. With the advent of social media, a signification portion of the public debate has shifted online~\cite{sch2015}, where news and views are disseminated through fragmented and ongoing conversations. However, these discussions are often highly polarized, making it challenging for a shared worldview to emerge~\cite{cinelli2021echo}.

The Internet has undoubtedly empowered the public by providing unprecedented opportunities for interaction and communication~\cite{edg2009}. However, digital platforms may also provide fertile ground for the proliferation of false or misleading information. Information disorder~\cite{wardle2017} can have severe consequences for society~\cite{zarocostas2020fight,cinelli2020covid}, especially when the affected topics are sensitive and emotive~\cite{larson2018biggest,bovet2019influence,betsch2015using}.
%such as those related to diseases~\cite{larson2018biggest}, politics~\cite{bovet2019influence} or vaccines~\cite{betsch2015using}.
As a result, a growing number of researchers, organizations, governments, and companies  are expressing concerns and devising countermeasures to mitigate the spread and impact of unsubstantiated content~\cite{twjack,allcott2017social,bovet2019influence,grinberg2019fake,avaaz_dis,EMM}.
%~\cite{twjack}. For example, in the recent past political elections have attracted lots of attention from scholars~\cite{allcott2017social,bovet2019influence,grinberg2019fake}, organizations~\cite{avaaz_dis}, and governments~\cite{EMM} investigating the presence and the countermeasure to the circulation of rumors.
One of the most widely used strategies for combating the dissemination of false or misleading information is to fact-check and rate the credibility of news sources. Websites such as MediaBiasFactCheck, AllSides, and Journalism Trust Initiative, as well as companies like NewsGuard, evaluate the accuracy and reliability of news sources based on various criteria. These evaluations often focus primarily on the verifiability of facts and the manner in which they are presented to audiences, with less attention given to editorial decisions such as the selection of events covered. This may be due to the additional research and analysis required to examine a news source's editorial line and coverage of all relevant events related to a given topic~\cite{shoe2009}. This imbalance is also reflected in the existing research on the subject. While there is a wealth of literature on the ideological biases of news sources~\cite{dellavigna2007fox,flaxman2016filter,groseclose2005measure} and their impact on public opinion~\cite{sharot2020people,bakshy2015exposure}, there is relatively less quantitative work on the topic of news gatekeeping~\cite{bourgeois2018selection}. Furthermore, there is a lack of understanding about the extent of selection bias in news publishing and its relationship to ideological bias.

In this article, we address both the bias in the selection of newsworthy events and the bias in the way the news is reported. \textit{Selection bias} refers to the tendency of a news outlet to choose certain events to cover while ignoring others. For instance, a news outlet may focus on adverse events related to vaccinations, while neglecting positive ones, thereby exhibiting a selection bias towards negative coverage. \textit{Narrative bias} refers to the way in which news events are framed and reported, potentially influencing the reader's perception and interpretation of the events. Together, these two forms of bias can significantly shape the way in which news is consumed and understood by the public.
%For example, suppose we consider an event (e.g., the results of a trial) related to vaccine effectiveness. This event can be presented by a news outlet reporting facts as they are (neutral narrative), exaggerating (pro-vax narrative), or denying (anti-vax narrative) the effectiveness of the vaccine. 
%This kind of bias is what debunkers and rating entities mainly take into account when assessing the reliability of a source.

Here we propose a novel approach to assessing both narrative and selection biases in news coverage. We investigate how these biases interact with each other and analyze their relationship with social engagement metrics. Additionally, we employ social network analysis to understand the connection between bias and news consumption.

To accurately quantify both narrative and selection biases, it is important to study news outlets' coverage of a widely-debated topic over an extended period of time. While narrative bias can be examined by considering a small sample of articles, selection bias can only be fully understood by analyzing a larger corpus of data over a longer time range. Indeed, selection bias reflects a news outlet's systematic preference for covering certain types of events, and such patterns can only be discerned by analyzing a significant amount of data over an extended time period.

As a case study, we focus on the topic of vaccines and immunization, which has been the subject of intense public debate for years, attracting a wide range of views and highly polarizing public opinion. We analyze the generation and dissemination of Italian news on vaccines over a six-year period from 2016 to 2021. During this time frame, the debate focused initially on the design, approval, and enforcement of the legislative framework on mandatory pediatric vaccinations (i.e., Law n.119 of July 31~\cite{legge119}), which was introduced in mid-2017 and fully implemented only in September 2019. Later on, the debate shifted to Covid-19 vaccines following the onset of the pandemic starting in January 2020.% (Italy was the first European country to be seriously hit by Covid-19 in early 2020).
Given the broad interest in this topic, its social significance, and the high degree of polarization, vaccines serve as an ideal subject to test our methodology.

We begin by conducting an analysis of the news, information, and views on vaccines produced by a diverse set of Italian news outlets that represent the most popular web sources with the largest reach on a wide range of social media platforms (Facebook, Instagram, Twitter, and YouTube). We rely on the assessment granted by independent fact-checking organizations, such as NewsGuard, Facta, Pagella Politica, and Butac, to differentiate between \textit{questionable} and \textit{reliable} sources. Moreover, we build an accurate machine learning model that is able to classify vaccine-related news based on the narrative being conveyed (\textit{anti-vax}, \textit{neutral}, \textit{pro-vax}) and the nature of the event being discussed (\textit{adverse}, \textit{neutral}, \textit{positive}). Using our dual classification, we are able to quantitatively measure and analyze how two distinct forms of bias (selection and narrative) shape the production of vaccine news. %These include narrative bias, which is represented by the different perspectives on vaccines (anti-vax, neutral, pro-vax) and selection bias, which refers to the types of events related to vaccines that are reported on (adverse, neutral, positive). 
Our approach allows us to separate these two forms of bias, compare them with the reliability classifications provided by independent third-party fact-checkers, and examine their relationship with engagement and consumption patterns. For example, we are able to determine whether a news source may not have a bias in the narrative conveyed but only reports on events of one type.

Hence, our contribution to the literature is three-fold:
i) First, we propose a methodology to measure not only narrative bias but also selection bias and compare them to the reliability classification provided by third-party. %As it will be pointed out in this paper, selection bias is a key element in the distortion of the public debate that often plagues modern communication.%, especially digital communication which is strongly linked to user engagement.
ii) Second, we analyze the relationship between these news-related biases and the online environment considering the spreading patterns and the engagement generated by the news sources. 
iii) Third, we provide a method to evaluate whether the ratings provided by external sources suffer from ideological bias. In other words, we pose the fundamental issue that can be epitomized in the Latin phrase of the Roman poet Juvenal: \emph{``Quis custodiet ipsos custodes?(i.e. who will guard the guardians?)''}

Our results show that the classification made by third-party entities is generally following the narrative bias dimension, but they fail to separate sources based on selection bias and are skewed towards a pro-vax narrative and a selection of positive events. We also show how highly biased sources are likely to create more engagement and we revealed how outlets with similar biases have a common audience.

\section*{Narrative Bias}
\begin{figure}
    \centering
    \includegraphics[width=0.49\textwidth]{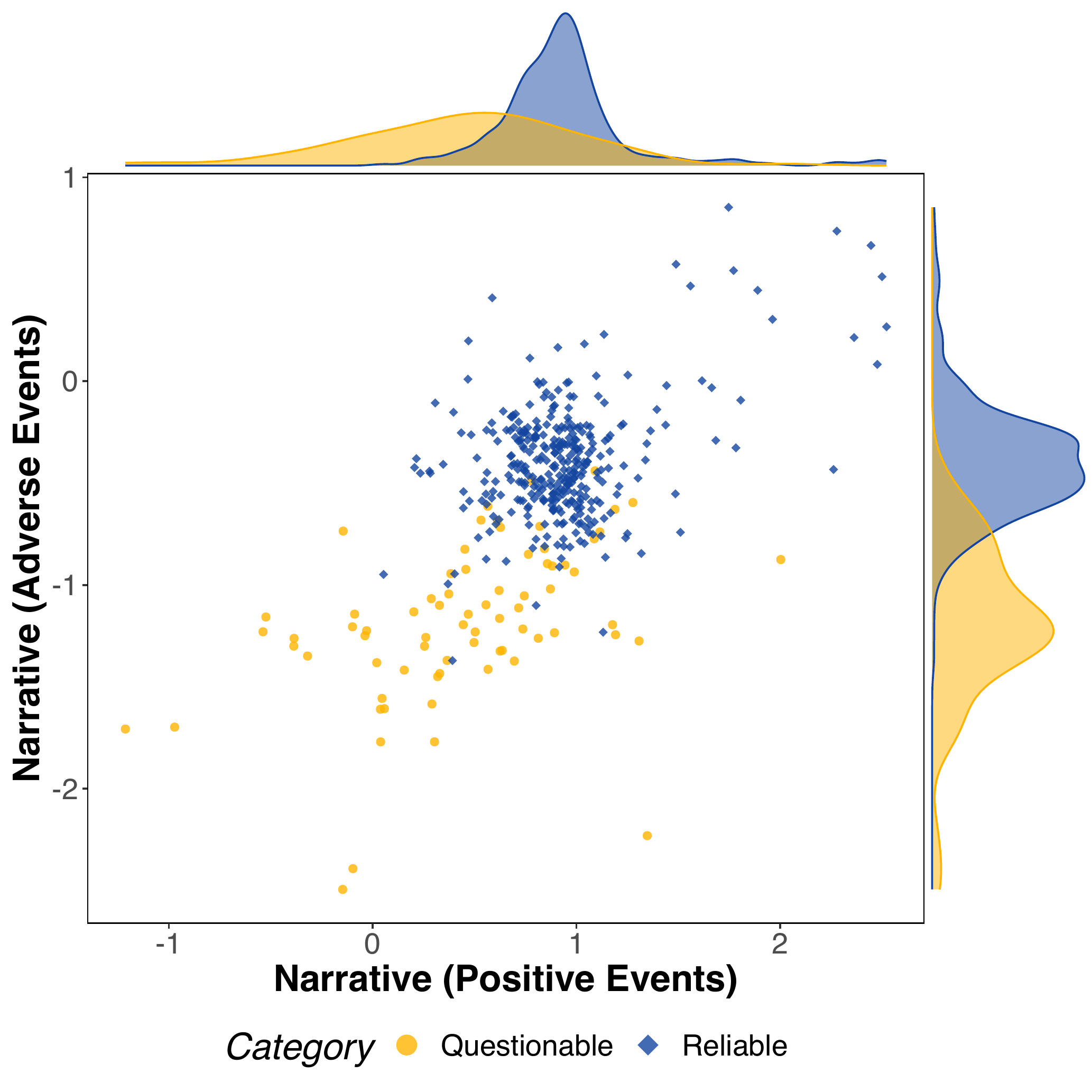}
    \caption{News outlets' narrative bias in reporting positive events against their estimated stance in reporting adverse events estimated by latent space model. Points are colored according to the classification retrieved from third-party data. Asymmetry in the axis values is due to the different framing strategies adopted when reporting on events of different nature (positive or negative).}

        \label{fig:estimates_static}
\end{figure}

We start by fitting a latent space model~\cite{barbera2015birds,hoff2002latent} to estimate a latent factor conveying information about the news outlets' narrative when dealing with adverse and positive events. This model takes as input the publishing behavior (i.e. information about the narratives and types of published articles) of news sources and maps each outlet on a latent dimension representing the narrative stance (see Materials and Methods section).
We set up our model so that the more negative (positive) the latent factor the stronger the anti-vax (pro-vax) narrative for both adverse and positive events. Figure~\ref{fig:estimates_static} shows the results of the latent factor estimation, where each dot represents the coordinates of a news outlet's narrative when dealing with positive ($x$-axis) or adverse ($y$-axis) events. Each dot is colored according to the reliability of the news outlet, as derived from third-party data (see section Materials and Methods for details).

As shown in Figure ~\ref{fig:estimates_static}, the estimated narrative bias aligns well with third-party classifications of news outlets as reliable or questionable. Additionally, the set of reliable outlets includes not only those with moderate positions, but also some of those with a strongly positive narrative. %shows an ideological leaning not only towards neutral narratives but also towards positive ones. 
The distinction between questionable and reliable outlets is further highlighted by their differing reporting styles. Questionable outlets tend to have a negative stance when reporting on both positive and adverse events, while reliable outlets have a milder position when reporting on adverse events and are more positive when reporting on positive events. This can be observed in the different modes of the marginal distributions. Additionally, we find that news outlets historically known for a strong conspiracy component are located in the bottom-left corner of the plot, while those with a strong pro-science position are located in the top-right corner.

\begin{figure*}
    \centering
    \includegraphics[width=0.89\textwidth]{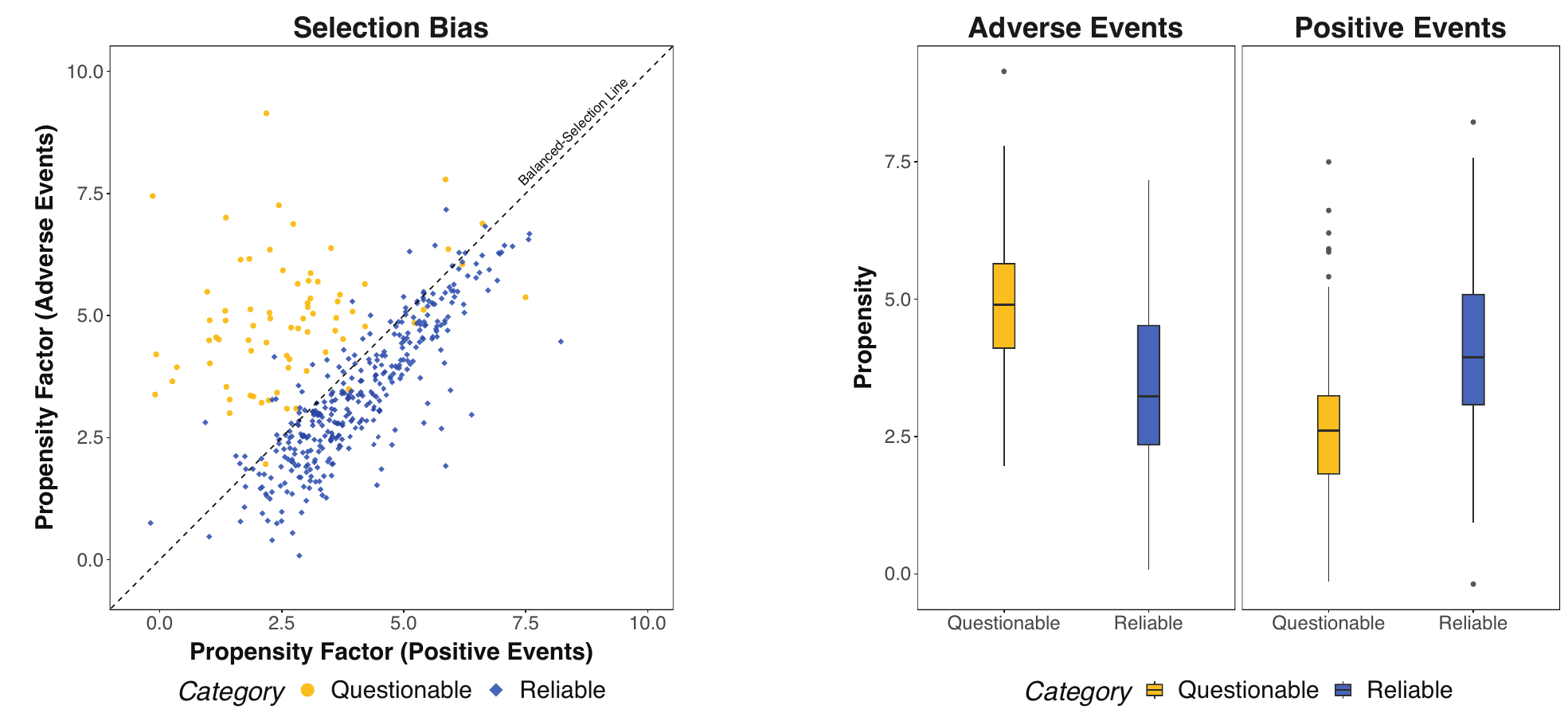}
    \caption{Estimated news outlets' propensity to report on positive events against negative events (left) and marginal distributions (right). The $45$-degree line represents the set of all points showing a balanced selection of news, i.e. equal propensity of reporting on positive and negative events. }
    \label{fig:estimates_propensity}
    
    % mettere da quache parte retta selection bias! 
\end{figure*}

\section*{Selection Bias}
%So far, we have analysed news outlets' narrative bias on vaccines. 
Selection bias, or gatekeeping, may also play a significant role in shaping news outlets' editorial lines and, as a result, public opinion.
To quantify this bias, i.e. the tendency of a news outlet to over- or under-discuss one type of newsworthy event, we use the intercepts of the estimated latent space models (as described in Methods). The intercept parameters $\alpha_{i, k= Adv}$ and $\alpha_{i, k = Pos}$, obtained as a by-product of the latent narrative estimation, represent the propensity of news outlet $i$ to report on adverse $(k= Adv)$ and positive events $(k= Pos)$, respectively. The higher $\alpha_{i k}$,  the stronger the propensity to report on that type of event.

To further understand the balance of a news outlet's reporting on positive and negative events, we have defined a \textit{Selection Index} (see Methods), which is calculated by measuring the distance between the point representing the news outlet and the diagonal in the adverse-positive propensity factor plane. The farther a point is from the diagonal, the more it favors one type of event in its articles. It is important to note that, unlike narrative bias, selection bias requires joint consideration of the propensity to write about positive and negative events in order to have a single index that provides information on the imbalance of reporting.
%Notice that here, differently from the case of narrative bias, it is possible to synthesize all the information in one index. While selection bias represents the propensity to report more on one type of event with respect to another one, narrative bias is instead the tendency to present a given type of event under a certain perspective, and cannot be separated by the events' type it belongs to. 

Figure \ref{fig:estimates_propensity} displays the propensity factors for positive and adverse events.  The 45-degree line represents a balance between the propensity to report on positive and negative events. %The 45-degree line indicates the set of all points showing a balanced propensity of publishing positive and negative events.
The figure shows that questionable news outlets have a strong propensity to report on adverse events and a weak propensity to report on positive events. The vast majority of reliable outlets show a balanced approach in their reporting, with a slight preference for positive events on average. Additionally, a small group of reliable outlets known for their strong pro-science position exhibit a strong propensity to report on positive events and a weak propensity to report on adverse events.

\section*{Biases and Engagement}
%%In the past sections, we analyzed the behaviour of news outlets and highlight the biases in the content they publish. However, online social media allow to analyse also how these content are perceived by the public.
Our next objective is to investigate the relationship between the strategies adopted by news outlets in terms of narrative and selection bias, and their level of user engagement. To do this, we must control for any scaling effects that may be present due to well-established news outlets having a larger audience and more resources for coverage. Therefore, we define an adjusted measure of engagement $E(s ; k; T)$:
\begin{equation*}
   E(s ; k; T)=\frac{I(s ;k; T)}{C(s ; k; T) \cdot F(s ; T)},
\end{equation*}
where $C(s; k; T )$ denotes the number of contents (i.e. articles) published by the news outlet $s \in S$ on events of type $k \in \{Adv, Neu, Pos \}$ in the time span $T$ and $I(s;  k; T )$ represents the corresponding number of user interactions (e.g. likes, shares, comments) received by $s$ on articles about events of type $k$ in the time span $T$, while $F(s; T)$ represents the average number of followers of the social media accounts of news outlet $s$ that were active during $T$.

\begin{figure}[t]
    \centering
    \includegraphics[width=0.5\textwidth]{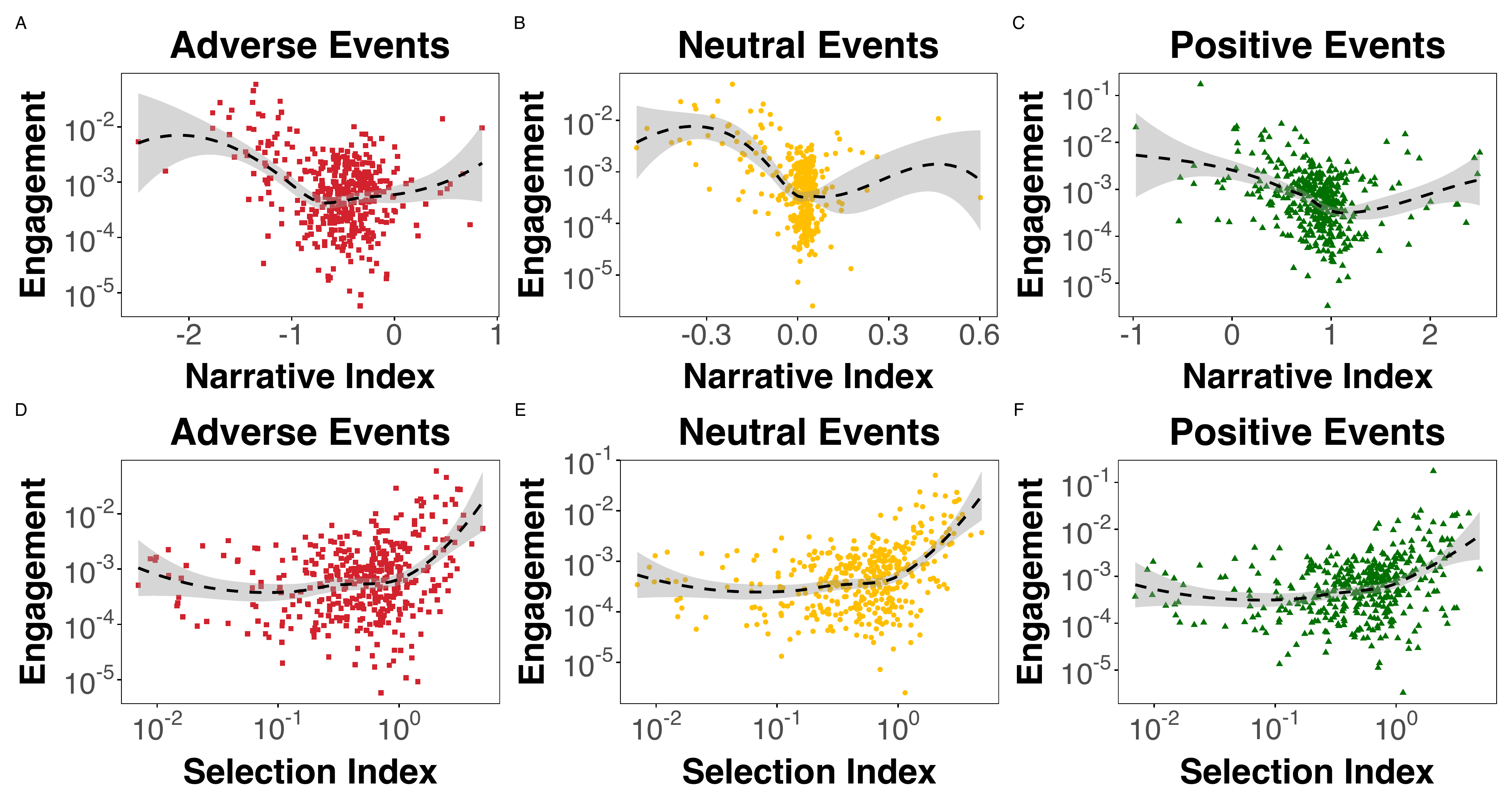}
    \caption{Scatter plots highlighting the non-linear relationship between Narrative bias and Engagement (top panels) and between Selection bias and Engagement (bottom panels) for what concerns Adverse (denoted by red squares), Neutral (yellow dots) and Positive events (green triangles). Adjusted Engagement is measured considering all the interactions to content(reactions, share and comments) and accounting for the size of each page.}
    \label{fig:eng_narrative}
\end{figure}

To gain insight into the relationship between engagement metrics and both narrative bias and selection bias proxies, we present scatter plots in Figure \ref{fig:eng_narrative}. The top panels display the relationship between the narrative bias factor and the engagement metric computed for adverse, neutral, and positive events, respectively. Similarly, the bottom panels show the relationship between the selection bias metric and the engagement metric. In both cases, the relationship appears strongly nonlinear. A U-shaped relationship appears to exist between engagement and narrative bias, with more extreme narratives seeming to be associated with higher engagement, while lower engagement is associated with moderate positions about the topic. However, this relationship appears to be mostly driven by the fact that questionable sources, characterized by a more negative outlook on the topic, are more successful in generating engagement. Also, a convex relationship seems to be in place when considering the selection bias metric. We further investigate these relationships through linear regression, which confirmed the existence of a convex relationship between narrative and selection biases and engagement(see SI). Overall, the plots suggest that being unbalanced in reporting news may increase engagement. 

\section*{Biases and News Consumption}

\begin{figure*}[t]
    \centering
    \includegraphics[width=0.99\textwidth]{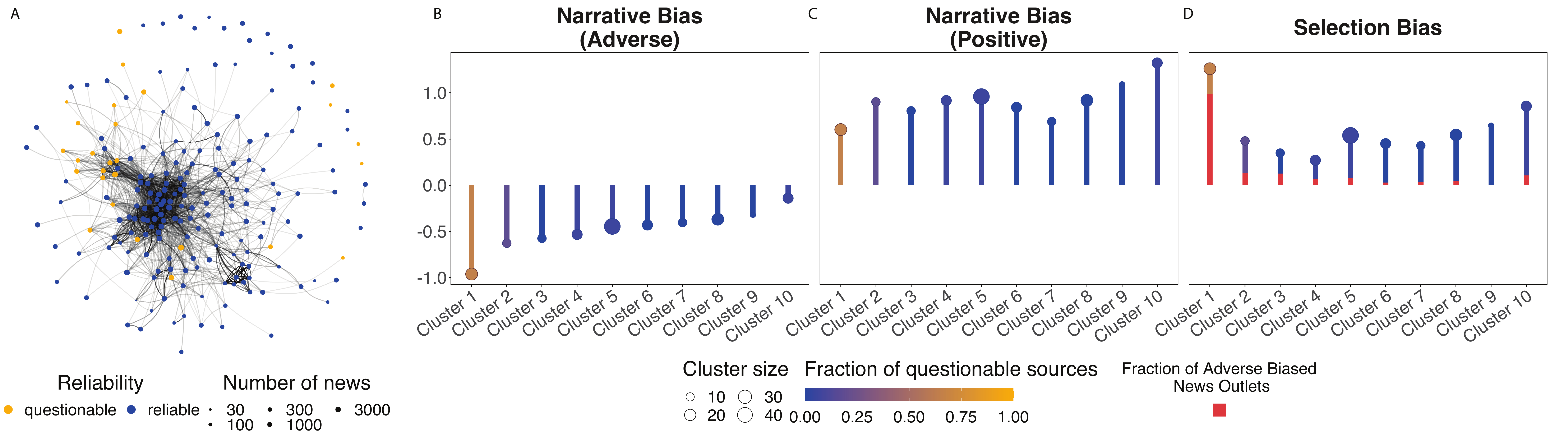}
    \caption{Outlets network of retweeters cosine similarity (panel a), Average narrative bias for adverse (panel b), and positive (panel c) events, and average selection bias (panel d) for clusters detected using Louvain algorithm. The reliability of news outlets is color-coded in the network and the fraction of questionable outlets in each cluster is color-coded, while the size of the dot is proportional to the cluster size.}
    \label{fig:cluster_hist}
\end{figure*}
%The analysis carried on in the previous section highlighted the relationship between news outlets biases and engagement.
A natural question is whether news outlets adopting similar publishing strategies also share a common audience. To examine this, we analyze the problem from the perspective of news consumption, using Twitter data on the vaccine debate from January 2020 to December 2021 to study the similarity in the audience of different news outlets (see SI). We define a metric based on cosine similarity on retweeters to quantify the connection between news outlets (see Materials and Methods). Intuitively, outlets sharing a high percentage of retweeters have a higher value of the similarity metric (close to 1), while outlets with only a few shared retweeters will have a low similarity (close to 0). Using this information, we build an undirected network in which nodes represent news outlets and weighted edges indicate the level of similarity. To highlight only the stronger connections, we discard edges with weights lower than the overall mean of the edges (see Materials and Methods).

The resulting network is visualized in Figure~\ref{fig:cluster_hist} and shows that reliable news outlets dominate the debate on Twitter. Moreover, reliable outlets form the core of the network, while questionable ones have a more peripheral role.
However, it is worth noting that there is no clear separation between questionable and reliable sources. This suggests that some users tend to retweet a set of reliable news outlets, while others have a mixed news diet, sharing both questionable and reliable sources. 
This is further supported by panels $B$,$C$, and $D$ of Figure~\ref{fig:cluster_hist}, where the percentage of questionable news outlets for each cluster detected using Louvain algorithm~\cite{blondel2008fast} is color-coded. 
Most of the clusters are primarily populated by reliable news outlets, with only one (Cluster 1) having more questionable (65\%) than reliable sources.
Furthermore, Figure~\ref{fig:cluster_hist} also reports the average value of narrative bias for adverse events (panel B) positive events (panel C), and selection bias (panel D), highlighting the differences between questionable and reliable clusters.

Indeed, the most questionable cluster (Cluster 1) has the lowest average narrative bias on adverse events, indicating that news outlets in this cluster tend to emphasize the magnitude of adverse events. At the same time, Cluster 1 also has the lowest narrative bias on positive events, meaning that its news outlets are likely to minimize the impact of positive events. Notably, this cluster has the highest value for selection bias, indicating that its news outlets do not cover adverse and positive events equally. On the other hand, Cluster 12 exhibits an opposite behavior. It has the highest narrative bias values for both adverse and positive events, indicating that its news outlets tend to minimize the importance of adverse events and exaggerate the importance of positive events. We also notice that this cluster has the second highest value of selection bias, implying that these news outlets do not cover both types of news equally.

In summary, the information presented suggests that news outlets in the most questionable and reliable clusters (Cluster 1 and Cluster 12, respectively) present events from opposing perspectives: the first one strongly endorses an anti-vax narrative, while the second one firmly promotes vaccination. Moreover, both exhibit a high selection bias, meaning that they tend to select only the type of events that align with their narrative.
To further verify this, we computed the fraction of news outlets that have a selection bias toward adverse events for each cluster, and showed it as a darker bar in panel D of Figure~\ref{fig:cluster_hist}. Intuitively, this is equivalent to the fraction of points that lie upon the 45-degree line in the left panel of Figure~\ref{fig:estimates_propensity}. Cluster 1 has the highest fraction of news outlets biased towards adverse events by far(78\%), while Cluster 12 has less than 13\%. Finally, a manual inspection of the members of the two more extreme clusters revealed that Cluster 12 is composed of news outlets widely recognized as pro-science, while Cluster 1 is populated by well-known anti-science and conspiracy-theory sources.

\section*{Conclusions}
In this study, we proposed a method to analyze the presence of biases in the selection, production, and dissemination of news. Our approach considered both the selection of events by news outlets (selection bias) and how they were presented (narrative bias) and consumed by citizens. We used machine learning techniques to classify the type (positive, neutral, or adverse) and narrative (pro-vax, neutral, anti-vax) of Italian vaccine-related events reported by news outlets. We used this information to fit a Bayesian model and quantified the two biases through latent variables. Moreover, using data from fact-checking agencies, we classified news outlets based on their reliability (reliable and questionable). Finally, we analyzed the relationship between news outlets' biases, citizen engagement, and news consumption.

Our results show that our method allows us for the quantification and assessment of the relevance of selection bias, which is much more difficult to assess than narrative bias and is often neglected in the quantitative literature. Our analysis also verifies the existence (or non-existence) of an ideological bias in fact-checking organizations whose role in public debate is to assess the quality of the information selected, disseminated, and discussed. By using data on the Italian debate about vaccines, our study highlights the role of selection bias in shaping public discussion on controversial issues. Furthermore, the method we have proposed is easily adaptable to other long-running and polarized public debates. All of these aspects - biases in the selection and narrative of newsworthy events, the impact of these on citizen reactions and news consumption patterns, and possible biases in the assignment of quality ratings by fact-checking organizations - are fundamental to the functioning of democratic societies. Understanding how these biases shape public opinion and discourse is crucial for ensuring an informed and engaged citizenry, which is vital for the health of any democracy.

\section*{Materials and Methods}
\subsection*{News Corpus Data}
We collected approximately 350K vaccine-related pieces of content published on Facebook, Instagram, Twitter, and YouTube from almost the entire universe of Italian news sources in the 6-year period from 01/01/2016 to 31/12/2021. The data collection process was carried out exclusively through the CrowdTangle API of Meta and the official APIs of Twitter and YouTube.
The selected news sources included a wide range of national/local newspapers, radio/TV channels, and online news outlets active in Italy during the aforementioned period, to ensure the most representative picture of both traditional and new media. Specifically, we selected 96 newspapers, 462 online-only news outlets, 89 TV channels, and 35 radio channels.
Then, using the 682 selected sources, we performed a keyword search for content that matched an exhaustive list of vaccine-related keywords, including general terms and vaccine brands/names (see SI for the complete list of keywords).

News sources were also assigned a binary label to distinguish between two categories: Questionable - i.e., a source producing mainly unverified or false content - and Reliable. The list of Questionable news sources was gathered by merging the lists from independent fact-checking organizations such as bufale.net, butac.it, facta.news, newsguardtech.com, and pagellapolitica.it. The remaining sources were labeled as Reliable. Table \ref{tab:breakdown} shows a breakdown of the dataset with the number of sources, contents, and corresponding user interactions (understood as the algebraic sum of all possible actions/reactions performed on the four platforms analyzed). Notice that the dataset is the same used in \cite{brugnoli2022vax} and for further details, including the complete list of news sources selected, we refer to that study.

\begin{table}[ht]
	\centering
	\def\arraystretch{1.1}
	\begin{tabular}{llll}
	\hline\hline
		Category & Sources & Contents & Interactions\\
		\hline
		Questionable & 161(23.6\%)& 44,547(12.6\%) & 10,898,774(11.4\%)\\
		%& (23.6\%) & (12.6\%) & (11.4\%)\\
		\hline
		Reliable & 521(76.4\%) & 308,983(87.4\%) & 84,332,137(88.6\%)\\
		%& (76.4\%) & (87.4\%) & (88.6\%)\\
		\hline
		Total & 682(100\%)  & 353,530(100\%) & 95,230,911(100\%) \\
		%& (100\%) & (100\%) & (100\%)\\
		\hline\hline
	\end{tabular}
	\caption{Breakdown of the dataset.}
	\label{tab:breakdown}
\end{table}

\subsection*{Twitter Data}
\label{sec:tw_data}
To analyze the similarity in news outlet audiences, we used Twitter data on the vaccination and Covid-19 vaccines debate. We collected all tweets made by the accounts we considered in our analysis that contained a keyword related to vaccination (see SI for the list of keywords used). We also retrieved all the retweets pointing to these tweets, obtaining a dataset of 23,908 tweets created by 315 news sources and 254,965 retweets created by 53,074 users. Notice that not all news outlets in our list had an active Twitter account.

\subsection*{Modeling Vaccine News Narrative and Event Type}
\label{sec:neural_models}
To classify the nature of the event reported (adverse, neutral, positive) and the narrative conveyed (anti-vax, neutral, or pro-vax) by vaccine-related content, we followed Google's pre-trained BERT multilingual cased model \cite{devlin2018bert}, which represents the state-of-the-art for semantic text representation in most languages \cite{abas2020deeplearn}, especially when data comes from social media \cite{tabinda2022transformers,vaswani2017attention}.
The narrative model is the same built in \cite{brugnoli2022vax} by training the BERT model on a manually annotated set of vaccine-related content, representing $\sim 10\%$ of the data gathered.
The sample was intentionally selected to contain anti- and pro-vax narratives.
Nonetheless, approximately half of the annotated data concerns neutral views.
To make the model more balanced between narrative classes and more confident with the local space around extreme values, augmented pieces of content \cite{ma2019nlpaug} were added to the sample by inserting words in a selection of data annotated as anti-vax or pro-vax through the contextual word embedding of BERT.
The same sample was here further annotated with the nature of the event reported and then used to fine-tune the BERT model for the corresponding classification task. The data to annotate was split among the authors to get $\sim 20\%$ overlap to compare the annotator agreement results with the model performance.
The augmented dataset was split into two parts to produce a dataset for training ($80\%$) and a dataset for evaluating ($20\%$) the model, by ensuring on both sets comparable class distributions with respect to narratives and events.
%To assure proper model evaluation, neither the annotated content used as a basis for the augmentation nor the augmented content were included in the evaluation set.
The annotation results with respect to the narrative and event for the training and evaluation sets are summarised in SI. 
The pre-trained BERT multilingual cased model consists of 12 stacked Transformer blocks with 12 attention heads each. We attached a linear layer with a softmax activation function at the output of these layers to serve as the classification layer. As input to the classifier, we took the representation of the special [CLS] token from the last layer of the language model. Both the narrative and event models were jointly trained end-to-end on the downstream task of three-class identification. We used the Adam optimizer with the learning rate of $5e-5$ and weight decay set to 0.01 for regularization. The models were trained for 4 epochs with batch size 64 through the HuggingFace Transformers library \cite{wolf2019huggingface}.
Statistics of the performances of the models are reported in the SI.

%Table \ref{tab:performance} reports the performance of the trained models compared with the inter-annotator agreement by using the same measure: accuracy (Acc) and the F1 score for individual classes, on both the training and the evaluation datasets. The confusion matrices for the evaluation set, used to compute all the scores of the annotator agreements and the model performance, are reported in SI.

\subsection*{A Latent Space Model for News Outlets' Stance}
The latent stance (in adverse, neutral, and positive events) was independently estimated by means of a latent space model \cite{hoff2002latent,barbera2015birds}. 
We modeled the number of news articles $y_{ijk}$ published by each news outlet $i \in \{1, \ldots, N\}$ within one of three categories, $M = 3$, namely anti-vax ($j = 1$), neutral ($j = 2$), pro-vax ($j = 3$), and for the subset of type-$k$ events (for  $k$ in \{Adverse, Neutral, Positive\}) via a Poisson distribution $y_{ijk} \sim \mathcal{P}ois(\lambda_{ijk})$ 
for which the log-intensity parameter is defined as
$log\lambda_{ijk} = \alpha_{ik}-||x_{ik}-z_{j}||$ where $||\cdot||$ denotes the euclidean distance between the stance of news outlet $i$, $x_{ik}$, and the ideal stance $z_j$, assumed to be such that $z_j \in \{-1, 0 , 1\}$, where $z_1 = -1$ is the stance associated to anti-vax, $z_2 = 0$ is the stance associated to neutral and $z_3 = 1$  is the stance associated to pro-vax.
We estimate the parameters of our model within a Bayesian setup by means of a Markov Chain Monte Carlo (MCMC) algorithm. On the one hand, the Bayesian estimation procedure allows for dealing with a complex model following a straightforward workflow. On the other hand, the Bayesian approach allows us to fully take into account the uncertainty associated with our estimates.
The prior specification for our set of parameters is the following: we assume a vague normal prior for the intercept parameter $\alpha_{ik}$, i.e. $\alpha_{ik} \sim \mathcal{N}(0, 15)$ for each $i$ and $k$ and an informative normal prior on the unit circle for the latent factor $x_{ik}$, i.e. $x_{ik} \sim \mathcal{N}(0, 1)$ for each $i$ and $k$. While we opted for a vague prior for the news-outlet-specific intercept parameter, we decided to set a more informative prior on the latent coordinates. This choice is a soft constraint that helps the identification of the latent coordinates, as suggested in \cite{barbera2015birds}.
The MCMC technique adopted in this case is a Metropolis-within-Gibbs \cite{robert1999monte} and is used to approximate the joint posterior of our model:
$$\pi(\boldsymbol{\theta}_k| Y_k) \propto  \left(\prod_{i = 1}^{N}\prod_{j = 1}^{M}f(y_{ijk}|\boldsymbol{\theta}_k)\right)\pi(\boldsymbol{\theta}_k),$$
where $\boldsymbol{\theta}_k = \{\boldsymbol{\alpha}_k,\mathbf{x}_k \}$ and $Y_k$ is the collection of $\{y_{11k}, \ldots, y_{NMk} \}$.

The algorithm implements the following steps:
\begin{enumerate}
    \item  Set $\boldsymbol{\alpha}_k^{(0)}$ and $\mathbf{x}_k^{(0)}$,
   \item  for each $h \in \{1, \ldots, H\}$
    \begin{enumerate}
    \setlength{\itemindent}{-1em}
    \item Draw $\alpha_{ik}^{(h)}$ from $\pi(\alpha_{ik}| \mathbf{x}_k, \boldsymbol{\alpha}_{-ik})$ via Random-Walk Metropolis-Hastings for each $i$;
    \item Draw  $x_{ik}^{(h)}$ from $\pi(x_{ik}| \mathbf{x}_{-ik}, \boldsymbol{\alpha}_k)$ via Random-Walk Metropolis-Hastings for each $i$;
    
    \end{enumerate}
\end{enumerate}
We notice that $H = 5,000$ is enough to obtain convergence after having discarded the first $1,000$ iterations as a burn-in. 

\subsection*{Selection Index}
To quantify how unbalanced a news outlet is in reporting on positive and negative events, we define the $SelectionIndex$, which is computed by measuring the distance between the location of any news outlet $i$ on the Propensity Factor plane and the $\theta$-degree line passing through the origin of the plane:
\begin{equation*}
SelectionIndex_i = \left|sin(\theta)\left(PF^{(Adv)}_i\right)-cos(\theta)\left(PF^{(Pos)}_i\right)\right|,
\end{equation*}
where the propensity factor of news outlet $i$ in reporting an event of type $k$, is the intercept parameter $\alpha_{i,k}$of the latent space model estimated on the set of type-k events, i.e $PF^{(k)}_i =\alpha_{i,k}$. We further assume that $\theta = \frac{\pi}{4}$, i.e., we consider a news outlet to be perfectly balanced if it shows an equal propensity to report on positive and adverse newsworthy events. 

\subsection*{News Outlets' Network}
We relied on the Twitter data described above and built an undirected weighted graph $G$ to quantify the similarity of news outlets in terms of audience.
We started creating a matrix $R$ with retweeters as rows and news outlets as a column. The entry $r_{i,j}$ of $R$ is the number of times user $i$ retweeted the news outlet $j$. We then compute the cosine similarity for each pair of columns to obtain the similarity measure for each pair of news outlets. Thus, the weight $w_{h,k}$ of the edge between node $h$ and $k$ in the graph $G$ is equal to:
\begin{equation*}
 w_{h,k}=\frac{r_h \cdot r_k}{||r_h|| ||r_k||},  
\end{equation*}
where $r_h$ and $r_k$ are the two column vectors of news outlets $i$ and $j$ respectively.
Notice that $w_h,k \in [0,1]$, since all the entries of the matrix are nonnegative.
Finally, we excluded all the 0-degree nodes and then all the edges with a weight below the mean weight of all edges, obtaining a graph with 206 nodes and 1,555 edges.
\vspace{0.5cm}
\\
\noindent
\textit{Authors Contribution}\\
All authors designed the research, A.G. and E.B. collected the data, A.G.,A.P.,E.B. performed the
analysis. All authors contributed to the manuscript.

\vspace{0.5cm}
\noindent
\textit{Competing Interests}\\
The authors declare no competing interests.

\bibliographystyle{IEEEtran}
\bibliography{references}
\end{document}